%% file: zac_submit.tex
\RequirePackage{fix-cm}
\documentclass[pdflatex,twocolumn,epjc3]{svjour3}
\smartqed  
\RequirePackage{graphicx}
\RequirePackage[colorlinks,citecolor=blue,urlcolor=blue,linkcolor=blue]{hyperref} 
 \usepackage{amsmath}
\usepackage{color}
\usepackage{multirow}
\input{gerda-abbreviations.tex}   
\journalname{Eur. Phys. J. C}
\begin{document}
\title{Improvement of the Energy Resolution via an\\
  Optimized Digital Signal Processing in \mbox{\protect\textsc GERDA} Phase~I} 
\titlerunning{ \mbox{{\textsc Gerda}} with Optimized Digital Signal Processing} 
%
\author{
 \mbox{\protect\textsc GERDA} collaboration\thanksref{corrauthor}\and ~\\[5mm] 
   M.~Agostini\thanksref{TUM} \and
   M.~Allardt\thanksref{DD} \and
   A.M.~Bakalyarov\thanksref{KU} \and
   M.~Balata\thanksref{ALNGS} \and
   I.~Barabanov\thanksref{INR} \and
   N.~Barros\thanksref{DD,alsoPO} \and
   L.~Baudis\thanksref{UZH} \and
   C.~Bauer\thanksref{HD} \and
   N.~Becerici-Schmidt\thanksref{MPIP} \and
   E.~Bellotti\thanksref{MIBF,MIBINFN}
% ...
}

\authorrunning{the \textsc{Gerda} collaboration}

\thankstext{corrauthor}{\emph{Correspondence}, email: gerda-eb@mpi-hd.mpg.de}
\thankstext{alsoPO}{\emph{present address:} Dept. of Physics and Astronomy,
  Univ. of Pennsylvania, Philadelphia, Pennsylvania, USA} 
\thankstext{alsoMIPT}{\emph{also at:} Moscow Inst. of Physics and Technology,
  Moscow, Russia} 
\thankstext{alsoIUN}{\emph{also at:} Int. Univ. for Nature, Society and
    Man ``Dubna'', Dubna, Russia} 
\institute{%
INFN Laboratori Nazionali del Gran Sasso, LNGS, and Gran Sasso Science
Institute, GSSI, Assergi, Italy\label{ALNGS} \and
INFN Laboratori Nazionali del Sud, Catania, Italy\label{CAT} \and
Institute of Physics, Jagiellonian University, Cracow, Poland\label{CR} \and
Institut f{\"u}r Kern- und Teilchenphysik, Technische Universit{\"a}t Dresden,
      Dresden, Germany\label{DD} \and
Joint Institute for Nuclear Research, Dubna, Russia\label{JINR} \and
Institute for Reference Materials and Measurements, Geel,
     Belgium\label{GEEL} \and
Max-Planck-Institut f{\"u}r Kernphysik, Heidelberg, Germany\label{HD} \and
Dipartimento di Fisica, Universit{\`a} Milano Bicocca,
     Milano, Italy\label{MIBF} \and
INFN Milano Bicocca, Milan, Italy\label{MIBINFN} \and
Dipartimento di Fisica, Universit{\`a} degli Studi di Milano e INFN Milano,
    Milan, Italy\label{MILUINFN} \and
Institute for Nuclear Research of the Russian Academy of Sciences,
    Moscow, Russia\label{INR} \and
Institute for Theoretical and Experimental Physics,
    Moscow, Russia\label{ITEP} \and
National Research Center ``Kurchatov Institute'', Moscow, Russia\label{KU} \and
Max-Planck-Institut f{\"ur} Physik, M{\"u}nchen, Germany\label{MPIP} \and
Physik Department and Excellence Cluster Universe,
    Technische  Universit{\"a}t M{\"u}nchen, M{\"u}nchen, Germany\label{TUM}
                     \and
Dipartimento di Fisica e Astronomia dell{`}Universit{\`a} di Padova,
    Padova, Italy\label{PDUNI} \and
INFN  Padova, Padova, Italy\label{PDINFN} \and
Physikalisches Institut, Eberhard Karls Universit{\"a}t T{\"u}bingen,
    T{\"u}bingen, Germany\label{TU} \and
Physik Institut der Universit{\"a}t Z{\"u}rich, Z{\"u}rich,
    Switzerland\label{UZH}
}
\date{Received: date / Revised version: date}
\maketitle
\begin{abstract}
 An optimized digital shaping filter has been developed for the
 \gerda\ experiment which searches for neutrinoless double beta decay in
 \gesix.  The \gerda\ Phase~I energy calibration data have been reprocessed
 and an average improvement of 0.3~keV in energy resolution (FWHM) at the
 $^{76}$Ge $Q$ value for $0\nu\beta\beta$~decay is obtained.  This is possible
 thanks to the enhanced low-frequency noise rejection of this Zero Area Cusp
 (ZAC) signal shaping filter.

\end{abstract}
%
\section{Introduction}
\label{intro}

\gerda~(GERmanium Detector Array)~\cite{gerdainstrument} searches for
neutrinoless double beta decay (\onbb\ decay) in \gess. The experiment is
located at the underground Gran Sasso National Laboratory (LNGS) of INFN,
Italy.  Crystals made from isotopically modified germanium with a fraction of
$\sim$86\,\% of \gess\ for a total mass of $\sim$20~kg are operated as source
and detector of the process.

Several extensions of the Standard Model of particle physics predict the
existence of \onbb\ decay, a process which violates lepton number conservation
by two units and which is possible if neutrinos have a Majorana mass
component.  \onbb\ decay is therefore of primary interest in the field of
neutrino physics.  Neglecting the nuclear recoil energy the energy released by
a \onbb\ event is shared by the two emitted electrons.  Both electrons are
stopped within $\sim$1~mm of germanium and thus all available energy is
deposited in a small region inside the detector.  Since distortions by
bremsstrahlung are expected to be small the \onbb\ decay signature is a
peak in the energy spectrum at the $Q$ value of the reaction, \qbb, amounting
to 2039~keV for \gess.  The most recent result of this process for \gess\ was
published by the \gerda~collaboration with a 90\,\% confidence level (CL)
limit on the \onbb\ half-life of
$T^{0\nu}_{1/2}>2.1\cdot10^{25}$~yr~\cite{gerdaresult}.

The sensitivity for detection of a possible \onbb\ decay signal depends on the
total efficiency $\varepsilon$ ($\simeq$75\,\% for \gerda\ Phase~I), the
enrichment fraction \fgesix\ and the isotopic mass $m_A$ of the considered
isotope, the total source mass $M$, the background level and the energy
resolution.  The expected number of signal events $n_S$\ for a given half-life
$T_{1/2}^{0\nu}$\ is~\cite{avignone}:
\begin{equation}\label{eq:signalcounts}
  n_S = \frac{1}{T_{1/2}^{0\nu}} \cdot \frac{\ln{2} \cdot N_A}{m_A} \cdot
                                 \fgesix\cdot \varepsilon \cdot M \cdot t
\end{equation}
where $N_A$~is the Avogadro number and $t$ the live time of the measurement.
The expected number of background events $n_B$ within an energy window $\Delta
E$\ is:
\begin{equation}\label{eq:bkgconts}
n_B = BI \cdot \Delta E \cdot M \cdot t
\end{equation}
with $BI$\ being the background index in \ctsper.  The size of $\Delta E$\ is
proportional to the energy resolution at \qbb, expressed as full width at
half-maximum (FWHM).  The energy resolution is of primary importance for the
enhancement of the sensitivity and the modeling of background sources.  If the
event waveforms are fully digitized with enough band width, the optimization
of energy resolution through a digital signal processing is possible.

A new energy reconstruction shaping filter leading to an improved energy
resolution has been developed (section~\ref{sec:zac}), that is denoted as Zero
Area Cusp (ZAC) filter.  The \gerda\ experiment (section~\ref{sec:gerda}), the
readout of the data (section~\ref{sec:dss}) and the signal processing
(section~\ref{sec:datacollection}) are described first. After the optimization
of the ZAC filter (section~\ref{sec:opti}) the Phase~I data have been
reprocessed (section~\ref{sec:results}).

\section{The \mbox{\protect{\textsc{GERDA}}} experiment}
\label{sec:gerda}

The design and the construction of \gerda\ were tailored to background
minimization.  The germanium detectors are mounted in low-mass ultra-pure
copper holders and are directly inserted in 64~m$^3$~of liquid argon (LAr)
acting as cooling medium and shield against external background radiation.
The argon cryostat is complemented by a water tank with 5~m diameter which
further shields from neutron and gamma backgrounds. It is instrumented with
photomultipliers to veto the cosmic muons by detecting \v{C}erenkov radiation.
A further muon veto is provided by plastic scintillators installed on the top
of the structure.  A detailed description of the experimental setup is
provided in Ref.~\cite{gerdainstrument}.

A first physics data collection, denoted as Phase~I, was carried out between
November 2011 and June 2013.  In Phase~I eight p-type semi-coaxial detectors
enriched in $^{76}$Ge from the Heidelberg-Moscow (\hdm)~\cite{HdM} and
IGEX~\cite{igex} experiments and five Broad Energy Germanium (BEGe) detectors
were used~\cite{sevenBEGe}.  Three coaxial detectors with natural isotopic
abundance from the Genius Test Facility (GTF) project~\cite{GTF1,GTF2} were
also installed.  In a second physics run (Phase~II) 30 BEGe detectors will be
operated in addition to the eight semi-coaxial together with instrumentation
to detect the LAr scintillation light to actively suppress
background~\cite{janickso,heisel,tipp2014}.

\subsection{Signal readout and shaping with germanium detectors}
\label{sec:dss}

\begin{figure}[t]
  \begin{center}
 \includegraphics[width=0.85\columnwidth]{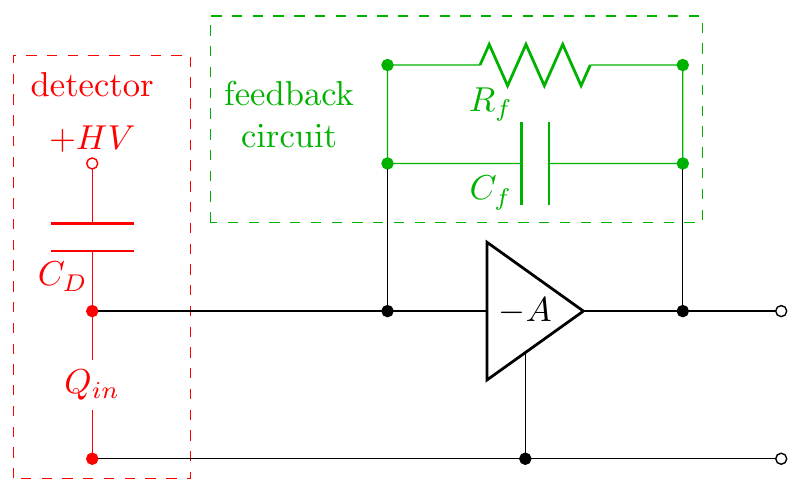}
  \caption{  \label{fig:frontend}       
    Typical readout scheme of a germanium detector with a charge sensitive
    preamplifier with open loop gain $A$. The detector with capacitance
    $C_D$ is operated with inverse bias voltage $HV$. The charge $Q_{in}$ is
    collected on the capacitor $C_f$ which then discharges because of the
    presence of the feedback resistor $R_f$.
}
\end{center}
\end{figure}

The typical readout of a germanium detector operated as a diode with inverse
bias voltage applied consists of a charge sensitive preamplifier whose output
wave form is either shaped and then processed by an analog to digital
converter or, as in \gerda, directly digitized by a flash analog to digital
converter (FADC).  Fig.~\ref{fig:frontend} presents the detector and the
charge sensitive preamplifier system consisting of a junction gate
field-effect transistor (JFET) coupled to a feedback circuit.  The capacitor
$C_f$ integrates the charge from the detector causing a steep change in
voltage at the preamplifier output.  In order not to saturate the dynamic
range of the preamplifier a feedback resistor $R_f$ is connected in parallel
to the capacitor to bring back the voltage to its baseline value.  The shape
of the preamplifier output pulse will then be characterized by a fast step,
with rise time of about 0.5--1.5~\mus\ corresponding to the charge collection
process followed by an exponential decay with time constant $\tau=R_fC_f$.
The values of $R_f$ and $C_f$ for the \gerda\ preamplifiers are 500~M$\Omega$
and $\sim$0.3~pF, respectively, for a $\tau$ of about 150~\mus.  A description
of the \gerda\ readout scheme is given in Ref.~\cite{gerdainstrument}.

\begin{figure}[t]
\begin{center}
\includegraphics[width=0.85\columnwidth]{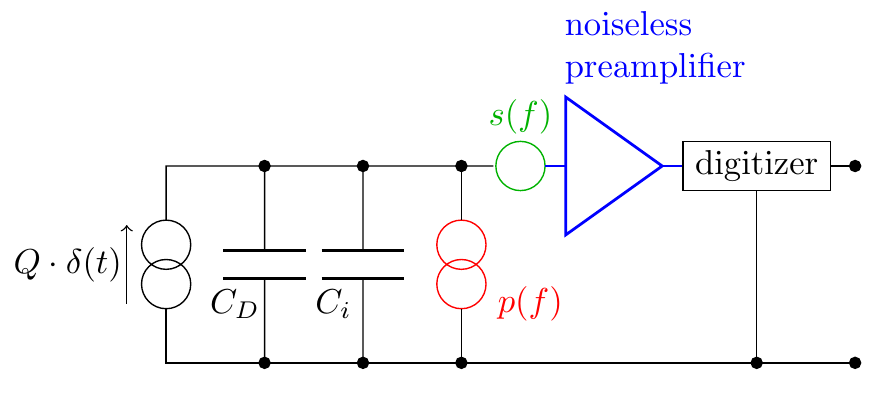}
   \caption{  \label{fig:noise}
        Signal and main noise sources in a germanium detector readout system.
        The trace recorded by the digitizer can be modeled as the output of a
        noiseless preamplifier, connected to a noiseless detector with
        capacitance $C_D$, a series voltage generator and a parallel current
        generator with spectral densities $s(f)$ and $p(f)$, respectively.
        $Q\cdot\delta(t)$ is the original current signal and $C_i$\ is the
        preamplifier input capacitance.
}
\end{center}
\end{figure}

Fig.~\ref{fig:noise} shows the signal and main intrinsic noise sources in the
detector and preamplifier system. The intrinsic equivalent noise charge (ENC)
for a given shaping time $\tau_s$ is given as:
\begin{equation}\label{eq:enc}
  ENC^2 = \alpha \frac{2kT}{g_m\tau_s}C_T^2 + \beta A_f C_T^2
                 + \gamma \Biggl( e(I_G+I_L) + \frac{2kT}{R_f} \Biggr) \tau_s
\end{equation}
where $g_m$ the JFET transconductance, $k$ is the Boltzmann constant, and
$T$~the operational temperature.  The constants $\alpha$, $\beta$ and $\gamma$
are of order 1 depending on the signal shaping filter
(c.f. Ref.~\cite{gattimanfredi}).  The series noise (first term) is
proportional to the total capacitance $C_T$ which is the sum of the detector
capacitance $C_D$, the feedback capacitance $C_f$~and the preamplifier input
capacitance $C_i$.  The second term represents the $1/f$~noise of the JFET
with amplitude $A_f$ and is also proportional to the total capacitance.  The
third term is the parallel noise generated by the detector leakage current
$I_L$, the gate current $I_G$ and the thermal noise of the feedback resistor
$R_f$.  The parallel noise is proportional to $\tau_s$ and the series noise to
its inverse while the $1/f$ noise is independent of $\tau_s$. Therefore, the
optimal shaping time is the one which minimizes the sum of the series and
parallel noise.  More detailed descriptions of the noise origin and its
treatment in germanium detectors can be found in Refs.~\cite{gattimanfredi}
and~\cite{radekanoise}.

In \gerda\ Phase~I an additional low-frequency disturbance comes from
microphonics related to mechanical vibrations of the long contacts (30--60~cm)
connecting the detectors to the preamplifiers.

\subsubsection{Digital Shaping}
\label{subsec:DigitalShaping}

\begin{figure}[]
\begin{center}
  \includegraphics[width=0.85\columnwidth]{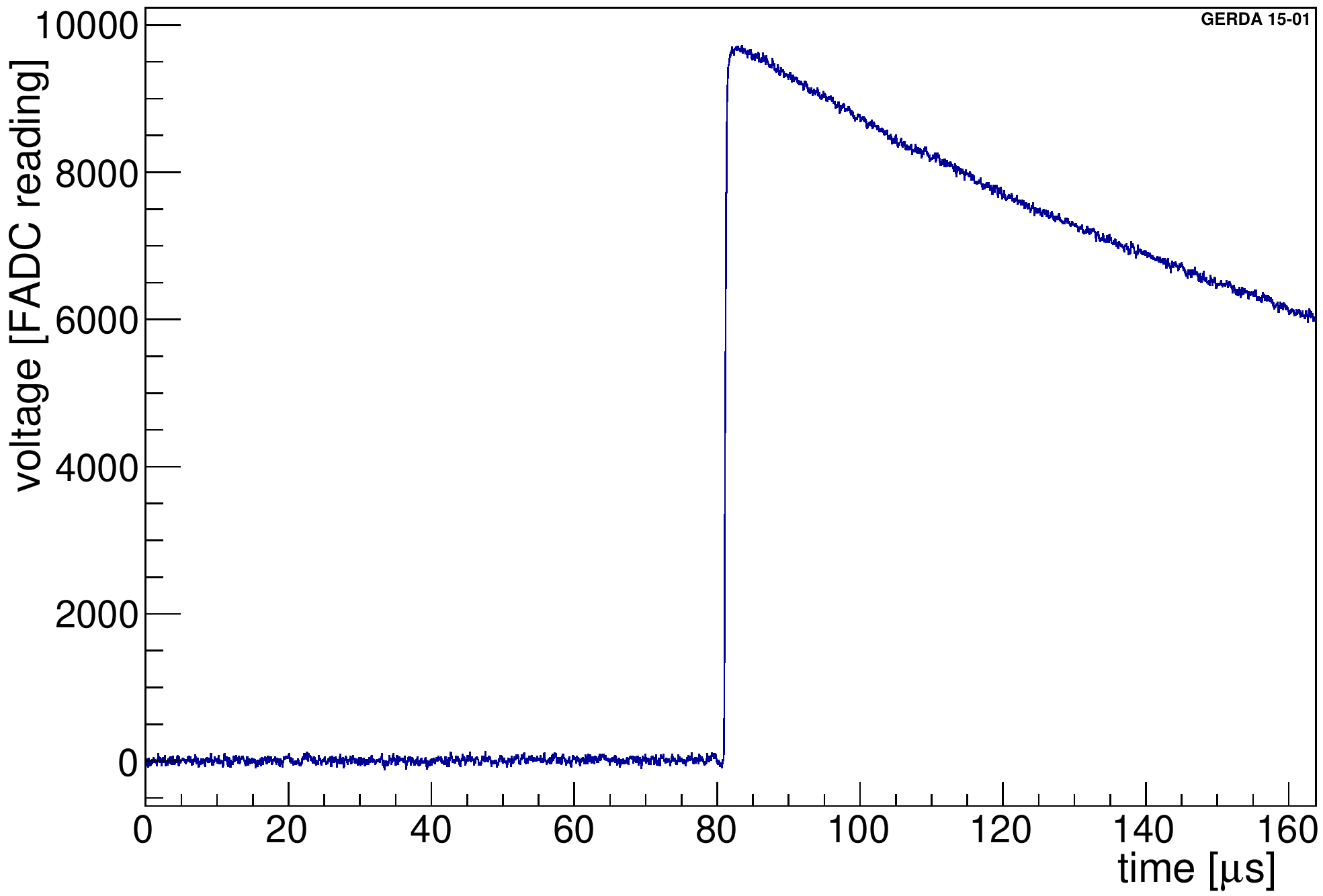}
  \caption{  \label{fig:originalwaveform}
          Typical wave form recorded in \gerda\ Phase~I. A $\sim$80~\mus\ long
          baseline is recorded before each signal. The exponential decay tail
          is from the discharge of the feedback capacitor.
}
\end{center}
\end{figure}

In \gerda\ Phase~I the signals were digitized with 14 bits precision and
100~MHz sampling frequency~\cite{gerdainstrument}.  16384 samples were
recorded per pulse (Fig.~\ref{fig:originalwaveform}).  After a $\sim$80~\mus\
long baseline the charge signal rises up with a $\sim$1~\mus\ rise time
followed by a $\sim$80~\mus~long exponential tail due to the discharge of the
feedback capacitor.

The energy estimation was performed by applying a shaping filter to the
digitized signal.  The advantages with respect to analog shaping are that a
large number of filters are available without restriction to the possible
settings of the analog shaping module and that raw data remain available for
further reprocessing.

\subsubsection{Energy Resolution}
\label{subsec:EnergyResolution}

The energy resolution of a germanium detector depends on the electronic noise,
on the charge production in the crystal and on the charge collection
properties of the diode and the shaping filter.  A hypothetical $\gamma$~line
at energy $E$ will have a $\Delta E$ (FWHM) expressed by:
\begin{equation}\label{eq:FWHMvsEnergy}
  \Delta E = 2.355\,\sqrt{\frac{\eta^2}{e^2}  ENC^2 +  \eta F\cdot E + c^2\,E^2 }
\end{equation}
where:
\begin{itemize}
\item $\eta$ is the average energy necessary to generate an electron-hole pair
  ($\eta=2.96$~eV in Ge) and $F$ is the Fano factor ($\sim0.1$~for
  Ge~\cite{lowe}).  This term contributes with about 1.8~keV at 2039~keV thus
  imposing a lower limit to the achievable $\Delta E$;
\item $c$ is a parameter related to the quality of the charge collection and
  integration.  An incomplete charge collection can be induced by charge
  recombination due to a too high impurity concentration or due to a too low
  bias voltage while a deficient integration of the collected charge can arise
  if a filter with a too short integration time is employed.  The same effect
  is obtained in all cases resulting in low-energy tails of the spectral
  peaks.  The parameter $c$ expresses therefore the amplitude of such tails.
  For the detectors used in \gerda\ Phase~I, the third term of
  Eq.~\ref{eq:FWHMvsEnergy} is usually one order of magnitude lower than the
  electronic and charge production terms for events with energy up to 3~MeV.
\end{itemize}

If the charge collection inefficiency is not dominant, the optimization of the
energy resolution depends almost exclusively on ENC, i.e. on the shaping
filter.  Given that ENC is independent of the energy, any $\gamma$ line with
sufficiently high statistics can be exploited for the optimization of the
shaping filter.

\subsection{Data collection and processing in \gerda}
\label{sec:datacollection}

Calibration data from the period Nov. 2011 -- May 2013
 were used to optimize the shaping filter.  The detectors considered
are ANG2--5 from the \hdm\ experiment, RG1--2 from IGEX and four of the five
BEGes (with names starting with ``GD'').  These are the same detectors used
for the \onbb\ decay analysis~\cite{gerdaresult}.  Since the electronic
disturbances could change as a function of the detector configuration in
\gerda, the calibration data were divided in four data sets as listed in
Table~\ref{tab:detconfiguration}.  In total 72 (45) calibration measurements
are available for the coaxial (BEGe) detectors.

\begin{table}[]
 \begin{center}
  \caption{  \label{tab:detconfiguration}
        Definition of data sets. The run ranges and active detectors are
        listed.
}
  \begin{tabular}{cll}
    \hline
    set & duration & detector configuration \\
    \hline
    A &09.11.11--22.05.12  & ANGs+RGs+GTFs \\
    B &02.06.12--15.06.12  & ANGs+RGs+GTF112 \\
    C &15.06.12--02.07.12  & ANGs+RGs+GTF112 \\
    D &08.07.12--21.05.13  & ANGs+RGs+GTF112+BEGes \\
    \hline
  \end{tabular}
 \end{center}
\end{table}

\subsubsection{Calibration of the energy spectrum}
\label{subsec:calibration}

\begin{figure*}[t]
\begin{center}
 \includegraphics[width=0.9\textwidth]{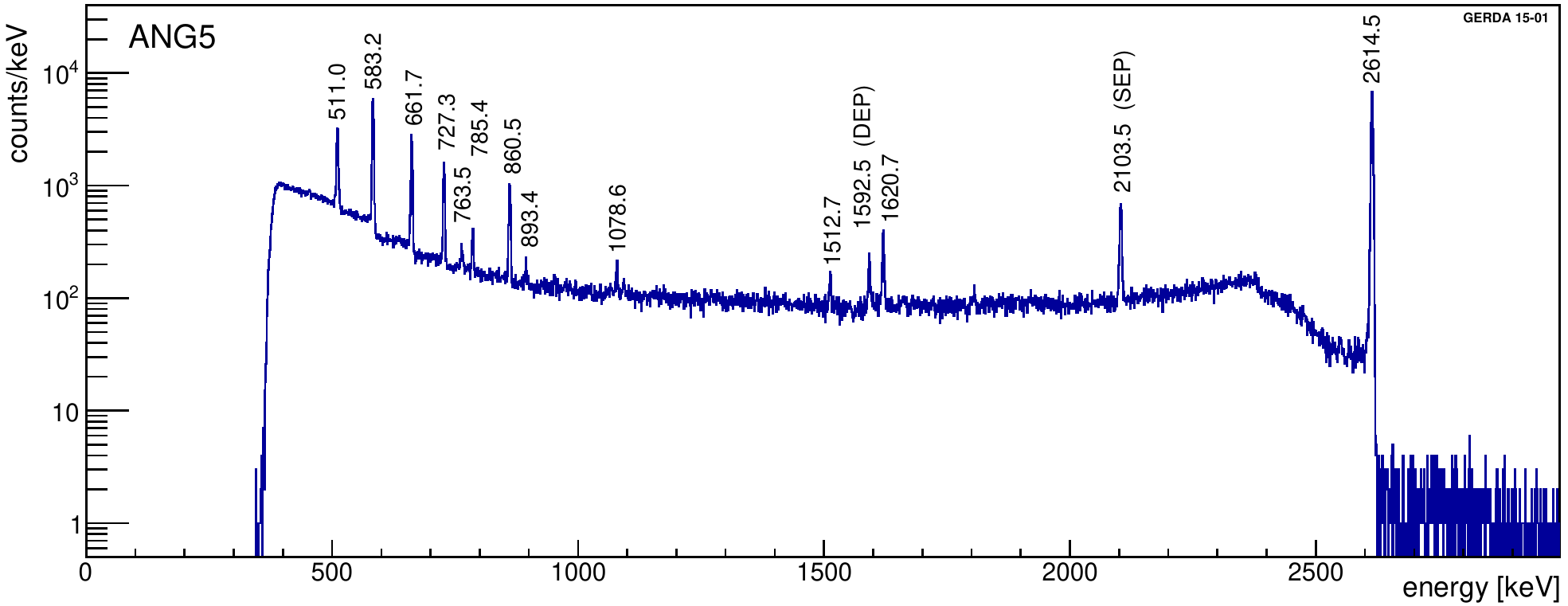}
  \caption{  \label{fig:CalibrationSpectrum_ANG5}
    A $^{228}$Th calibration spectrum recorded by ANG5.
    The threshold is set to $\sim$400~keV.
}
\end{center}
\end{figure*}

The calibrations were performed by inserting up to three $^{228}$Th sources in
proximity of the detectors~\cite{tarka,francis}.  The total activity of the
sources was about 40~kBq at the beginning of Phase~I. The duration of the
measurements was between one and two hours.  The energy threshold for the
calibrations is $\sim$400~keV to reduce disk usage.  At least ten peaks with
energies between 0.5 and 2.6~MeV are visible in the recorded spectra
(Fig.~\ref{fig:CalibrationSpectrum_ANG5}).  While all peaks are exploited for
the calibration of the energy scale, only the full energy peaks (FEP) are used
in the fit of the FWHM as function of energy.  This is necessary because the
single escape peak (SEP), the double escape peak (DEP) and the 511.0~keV line
are Doppler broadened.

Given the large number of calibration spectra to be analyzed, a fully
automatized routine was developed and used throughout Phase~I.  The main steps
of the procedure are:
\begin{itemize}
\item rejection of events which might decrease the precision of the
  calibration; e.g., coincidences between detectors, wave forms with
  superimposed events (pile-up events);
\item search and identification of the peaks;
\item fit of the peaks and automatic adjustment of the fitting function
  according to the number of events in the peak and the peak shape;
\item extraction of the calibration curve;
\item fit of the FWHM as a function of energy.
\end{itemize}

\subsubsection{Signal processing}
\label{subsec:signalprocessing}

The signal processing of \gerda\ Phase~I data was performed through an offline
analysis of the digitized wave forms with the software tool
\gelatio~\cite{gelatio}.  The standard energy reconstruction algorithm is a
digital pseudo-Gaussian filter consisting of:
\begin{itemize}
\item a delayed differentiation of the sampled trace
  \begin{equation}
    x_0[t] \rightarrow x_1[t] = x_0[t]-x_0[t-\delta]
  \end{equation}
  where $x_0[t]$ is the signal height at time $t$ and $\delta$ was chosen to be
  5~\mus;
\item the iteration of 25 moving average (MA) operations: 
  \begin{equation}
    x_i[t] \rightarrow x_{i+1}[t] = \frac{1}{\delta}\sum_{t' = t - \delta}^t x_i[t']
    \quad\quad i=1,\dots,25
  \end{equation}
\end{itemize}
The energy is given by the height of the output signal whose shape is close to
a Gaussian.

This pseudo-Gaussian shaping is a high-pass filter followed by $n$ low-pass
filters.  The resolution obtained with the pseudo-Gaussian shaping is very
close to optimal if the detectors are operated in conditions where the $1/f$
noise is negligible~\cite{gattimanfredi}.  This is not the case for
\gerda\ Phase~I where the preamplifiers had to be placed at a distance of
30--60~cm from the crystals due to the low background requirements.  The
diodes and the pre-amplification chain were connected by copper stripes.
Hence, a significant low-frequency noise is present for some of the
\gerda\ Phase~I detectors.

As described in Sec.~\ref{sec:dss}, the ENC depends on the properties of the
detector, of the preamplifier and of the connection between them.  In \gerda\
the diodes have different geometries and impurity concentrations resulting in
different capacitances $C_D$~and different $I_L$.  In addition, the
non-standard connections between the detectors and the preamplifiers result in
different input capacitances ($C_i$).  It is therefore preferable to adapt the
form and the parameters of the shaping filter to each detector separately.

\section{ZAC: a novel filter for enhanced energy resolution}
\label{sec:zac}

Several methods have been developed to obtain the optimum digital shaping for
a given experimental
setup~\cite{gattimanfredi,radekanoise,deighton,gattisampietro}.  For
series and parallel noise and with infinitely long wave forms it
can be proven~\cite{deighton} that the optimum shaping filter for energy
estimation of a $\delta$-like signal is an infinite cusp with the sides of the
form $\exp{(t/\tau_s)}$ where $\tau_s$~is the reciprocal of the corner
frequency; \emph{i.e.}, the frequency at which the contribution of the series
and parallel noise of the referred input become equal.  When dealing with wave
forms of finite length, a modified cusp is obtained in which the two sides
have the form of a {\it sinh}-curve.  If low-frequency noise and disturbances
are also present, the energy resolution is optimized using filters with total
area equal to zero~\cite{geracigatti}.  In addition, the low-frequency
baseline fluctuations (e.g. due to microphonics) are well subtracted by
filters with parabolic shape~\cite{geracirech}.  The best energy resolution
for \gerda\ is achieved if a finite-length cusp-like filter with zero total
area is employed.  This can be obtained by subtracting two parabolas from the
sides of the cusp filter keeping the area under the parabolas equal to that
underlying the cusp.

In reality the detector output current is not a pure $\delta$-function, but
has a width of approximately 1~\mus.  If a cusp filter is used, this leads to
the effect of a ballistic deficit~\cite{radekatrap,goulding} and consequently to
the presence of low-energy tails in the spectral peaks.  This can be remedied
by inserting a flat-top in the central part of the cusp with a width equal to
almost the maximum length of the charge collection in the diode.  The
resulting filter is a Zero-Area finite-length Cusp filter with central
flat-top that will be referred as ZAC from here on.

\begin{figure}[t]
\begin{center}
  \includegraphics[width=0.9\columnwidth]{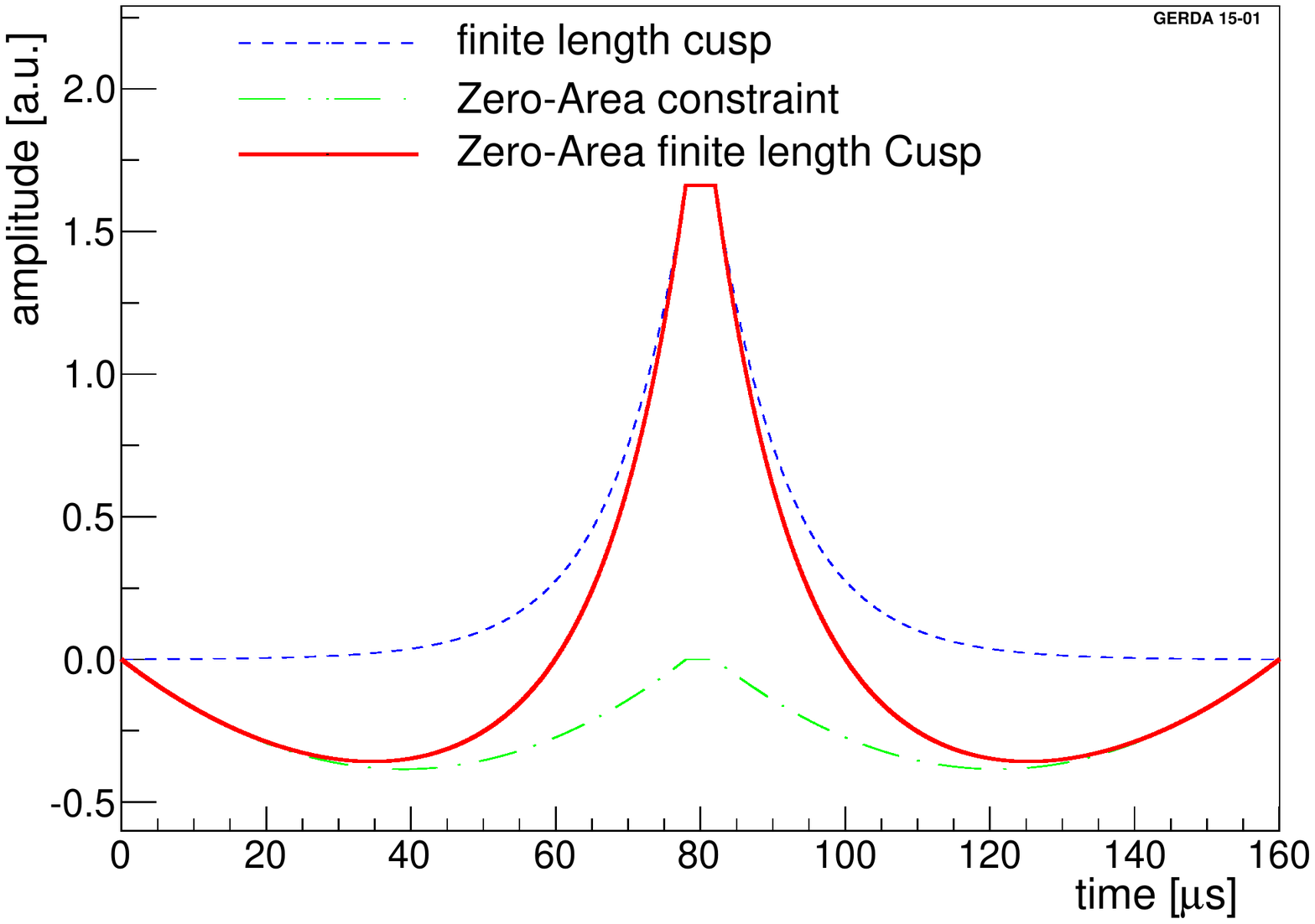}
  \caption{  \label{fig:ZAC}
            Amplitude versus time for the ZAC filter (red full line). It is
            composed of the finite-length cusp (blue dashed) from which two
            parabolas are subtracted on the cusp sides (green dash-dotted).
}
\end{center}
\end{figure}

The ZAC filter was implemented as:
\begin{multline}\label{eq:zac}
  ZAC(t) =\\
  \begin{cases}
    \sinh \Bigl( \frac{t}{\tau_s} \Bigr) + A\cdot \Bigl[ \bigl(t-\frac{L}{2}\bigr)^2 - \bigl(\frac{L}{2}\bigr)^2 \Bigr]  & \\ \hfill 0 < t < L\\ 
     \sinh \Bigl( \frac{L}{\tau_s} \Bigr) & \\ \hfill L < t < L + FT \\[2mm]
     \sinh \Bigl( \frac{2L+FT-t}{\tau_s} \Bigr) + A\cdot \Bigl[ \bigl(\frac{3}{2}L+FT-t\bigr)^2 
  -\bigl(\frac{L}{2}\bigr)^2 \Bigr]  & \\[2mm] \hfill L+FT < t < 2L+FT
  \end{cases}
\end{multline}
where $\tau_s$ is the equivalent of the shaping time for an analog shaping
filter, $2L$ is the length of the cusp filter and $FT$ is that of the flat-top
and where the constant $A$ is chosen such that the total integral is zero. The
numerical expression of the ZAC filter is obtained through the substitution
$t\rightarrow\Delta t\cdot i$ where $\Delta t$\ is the sampling time and
$i$\ the sample index; the maximum number of samples in the ZAC filter is
$n_{ZAC}$. A graphical representation of the ZAC filter construction is
provided in Fig.~\ref{fig:ZAC}.

\begin{figure}[t]
\begin{center}
  \includegraphics[width=0.85\columnwidth]{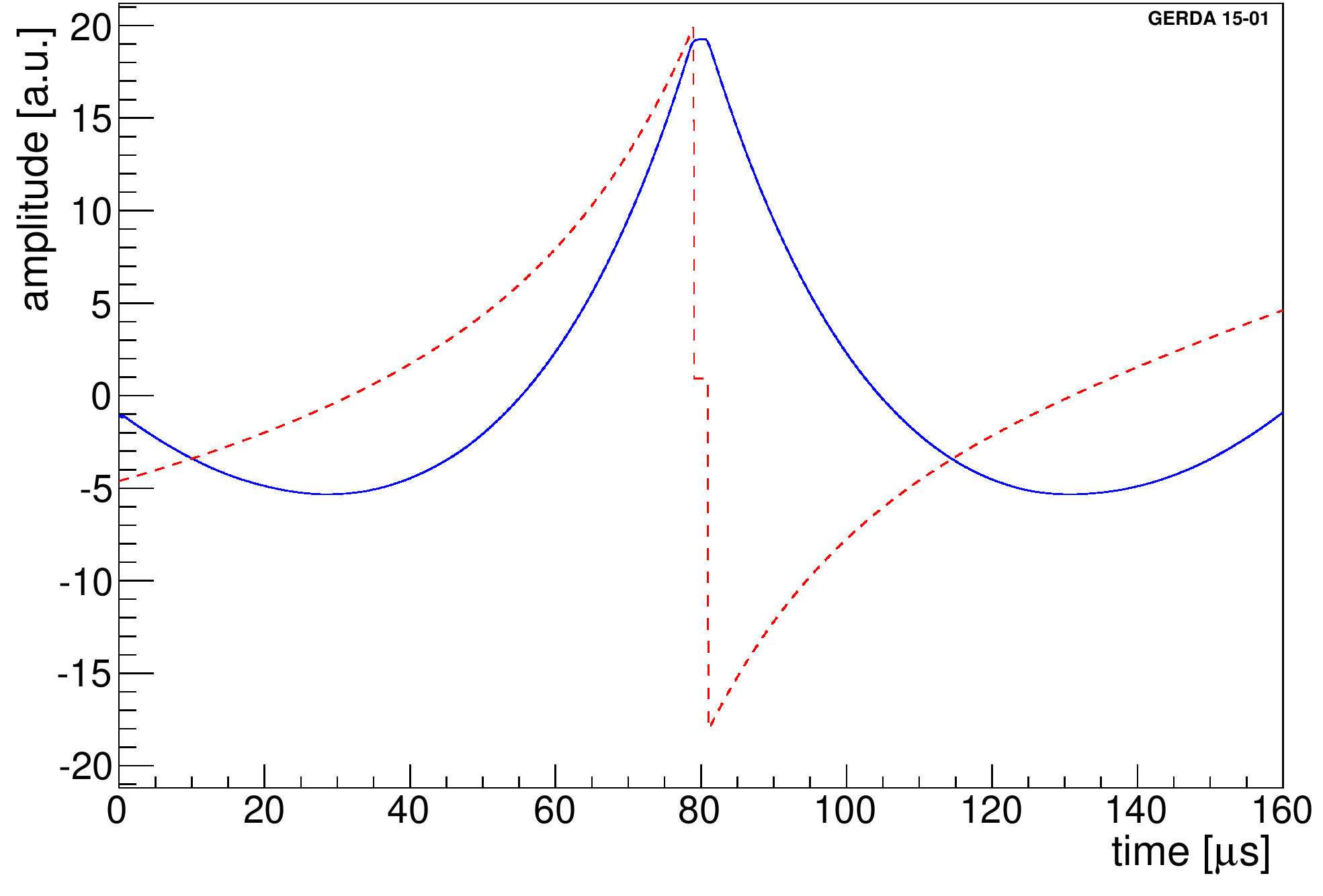}
  \caption{  \label{fig:ZACdeconvolved}
     The ZAC filter after the convolution with the inverse preamplifier
     response function (red dashed) and the wave form of
     Fig.~\ref{fig:originalwaveform} after the convolution with it (full blue).
}
\end{center}
\end{figure}

Before proceeding with the shaping the original current pulse has to be
reconstructed from the preamplifier output wave form
(Fig.~\ref{fig:originalwaveform}).  This is performed via a deconvolution of
the preamplifier response function, an exponential curve with decay time
$\tau=R_fC_f$.  Specifically, it is implemented as the convolution with the
filter consisting of 2 elements,
$f_{\tau}=\Bigl[1,-\exp{\Bigl(-\frac{\Delta t}{\tau}\Bigr)}\Bigr]$.
No correction for the finite band width of the electronics was implemented.
Since the convolution operation is commutative, the convolution between the
ZAC filter and the inverse preamplifier response function $f_{\tau}$\ can be
performed once for all:
\begin{eqnarray}\label{eq:filterconvolution}  \nonumber
  FF[i] &=& ZAC[i] \cdot \Bigl(-e^{-\frac{\Delta t}{\tau}}\Bigr)
           + ZAC[i+1] \cdot 1\\
 && \hspace*{35mm} i=1,...,n_{ZAC}-1
\end{eqnarray}
The final filter ($FF$) obtained is shown in red in
Fig.~\ref{fig:ZACdeconvolved}.  A convolution of each individual signal trace
$x$\ with  $FF$ is then performed:
\begin{multline}\label{eq:finalconvolution}
  y[i] = \sum_{k=i}^{i+n_{ZAC}-2} x[k] \cdot FF[i+n_{ZAC}-1-k] \\
  i=1,...,n_x-n_{ZAC}+2
\end{multline}
$n_x$ is the number of samples in the trace.  Typically $n_x$ is set to 16384
and $n_{ZAC}$ ranges from 16060 to 16120. The output~$y$ for the trace of
Fig.~\ref{fig:originalwaveform} is shown as blue full line in
Fig.~\ref{fig:ZACdeconvolved}.  The energy $E$ is then estimated as the
maximum of this convoluted signal $y$.

\section{Optimization of the ZAC filter on calibration data}
\label{sec:opti}

The optimization of the ZAC filter using the Phase~I calibration data was
performed separately for each detector.  The first and the last calibration
run of each period were selected (Table~\ref{tab:detconfiguration}).  Given
their longer duration one more run taken in the middle of the period was used
for data sets A and D as well.  It is expected that no change was present in
the electronic noise within the same data set.  In this case the filter
parameters giving the best energy resolution should be the constant for each
data set.

The filter optimization was performed on the FEP of $^{208}$Tl, i.e. the
2614.5~keV line.  Quality cuts were applied prior to the energy reconstruction
that was performed only on the surviving events.  The energy spectrum was
reconstructed with different values of the four filter parameters $L$, $FT$,
$\tau_s$ and $\tau$.  In particular:
\begin{itemize}
\item the total filter length $2L+FT$ was varied for only one calibration run
  between 120 and 163~\mus.  As expected~\cite{deighton} the best energy
  resolution was obtained for the longest possible filter.  Given the
  variability of the trigger time within a 2~\mus\ range the maximum of the
  shaped filter can be at one of the extremes of the wave form when the
  maximum filter length of 163~\mus\ is used leading to a wrong energy
  estimation.  This effect completely disappears if the filter is shortened by
  2~\mus.  Hence, the optimization was performed with $\sim$161~\mus\ long
  filters;
\item the optimal length of $FT$ is related to the charge collection time in
  the detector.  For coaxial detectors this is typically between 0.6 and
  1~\mus\ depending on the electric field configuration in the detector and on
  the location of the energy deposition.  For BEGes it is slightly longer due
  to the slower charge drift.  The value of $FT$ was therefore varied between
  $0.5$~and 1.5~\mus\ in 120~ns steps;
\item the optimal filter shaping time $\tau_s$ depends on the electronic noise
  spectrum as described in Sec.~\ref{sec:dss}. Typically, $\tau_s$ is of order
  of 10~\mus.  The optimization was therefore performed with values of
  $\tau_s$\ between 3 and 30~\mus\ in steps of 1~\mus. Since the optimal
  $\tau_s$\ was not infinite, the noise present in Phase~I data had a non
  negligible parallel component;
\item the value of $\tau$ can in principle be calculated knowing the feedback
  resistance and capacitance.  In reality $\tau$ is modified by the presence
  of parasitic capacitance in the front-end electronics.  Moreover, given the
  presence of long cables a signal deformation can arise.  Therefore, $\tau$
  is normally estimated by fitting the pulse decay tail.
  This was not possible due to the presence of more than one exponential.
  Therefore $\tau$ was varied between 100 and 300~\mus\ with 5~\mus\ step size.
\end{itemize}

\begin{table}[b]
 \begin{center}
  \caption{  \label{tab:optimizedparameters}
    Optimized parameters of the ZAC filter for period~D.
    While the filter length $2L$ is equal for all the detectors
    $FT$ varies between 0.6 and 1.2~\mus\ according to the charge
    collection properties of each diode.}
  \begin{tabular}{ccccc}
    \hline
    detector & 2L [\mus] & FT [ns] & $\tau_s$ [\mus] & $\tau$~[\mus] \\ 
    \hline
    ANG2  & 160 & 600  & \ 9  & 190 \\
    ANG3  & 160 & 840  & 16   & 220 \\
    ANG4  & 160 & 720  & 13   & 250 \\
    ANG5  & 160 & 960  & 17   & 170 \\
    RG1   & 160 & 720  & 12   & 210 \\
    RG2   & 160 & 680  & \ 8  & 240 \\
    GD32B & 160 & 1080 & 13   & 220 \\
    GD32C & 160 & 960  & 16   & 170 \\
    GD32D & 160 & 840  & \ \ 15.5 & 170 \\
    GD35B & 160 & 1200 & 17   & 135 \\
    \hline
  \end{tabular}
 \end{center}
\end{table}

The peak at 2614.5~keV was fitted with the function~\cite{phillipsmarlow} for
each combination of the filter parameters:
\begin{eqnarray} 
\label{eq:FEPfit}  \nonumber
  f(E) &=& A\exp{\Biggl( -\frac{(E-\mu)^2}{2\sigma^2} \Biggr)} + B +
  \frac{C}{2} \mbox{erfc} \Biggl( \frac{E-\mu}{\sqrt{2}\cdot\sigma} \Biggr)  \\
  &&+\frac{D}{2} \exp{\Biggl( \frac{E-\mu}{\delta} \Biggr)} \mbox{erfc}
    \Biggl( \frac{E-\mu}{\sqrt{2}\cdot\sigma} 
   + \frac{\sigma}{\sqrt{2}\cdot\delta} \Biggr)
\end{eqnarray} 
corresponding to a Gaussian peak with a low-energy tail (last term) sitting on
flat background and on a step-like function (third term) which describes the
continuum on the left side of the peak. The FWHM was obtained from the fitting
function after the subtraction of the flat and step-like background
components.  The energy resolutions resulting from different parameters of the
ZAC filter were compared and the parameters leading to a minimal FWHM were
chosen for the full reprocessing of the data. For the detectors of the
\onbb\ analysis the optimal parameters of the ZAC filter for period~D are
reported in Table~\ref{tab:optimizedparameters} as an example.

\section{Results}
\label{sec:results}

The parameter optimization for the ZAC filter provided results in agreement
with expectations: for each detector the optimal filter parameters are stable
within the same data set, but they can vary for those detectors that changed
configuration in time.  This confirms the dependence of the microphonic
disturbances on the cable routing.  Hence, all Phase~I calibration and physics
data were reprocessed with the optimal parameters of the ZAC filter.

\begin{figure}[t]
\begin{center}
  \includegraphics[width=0.9\columnwidth]{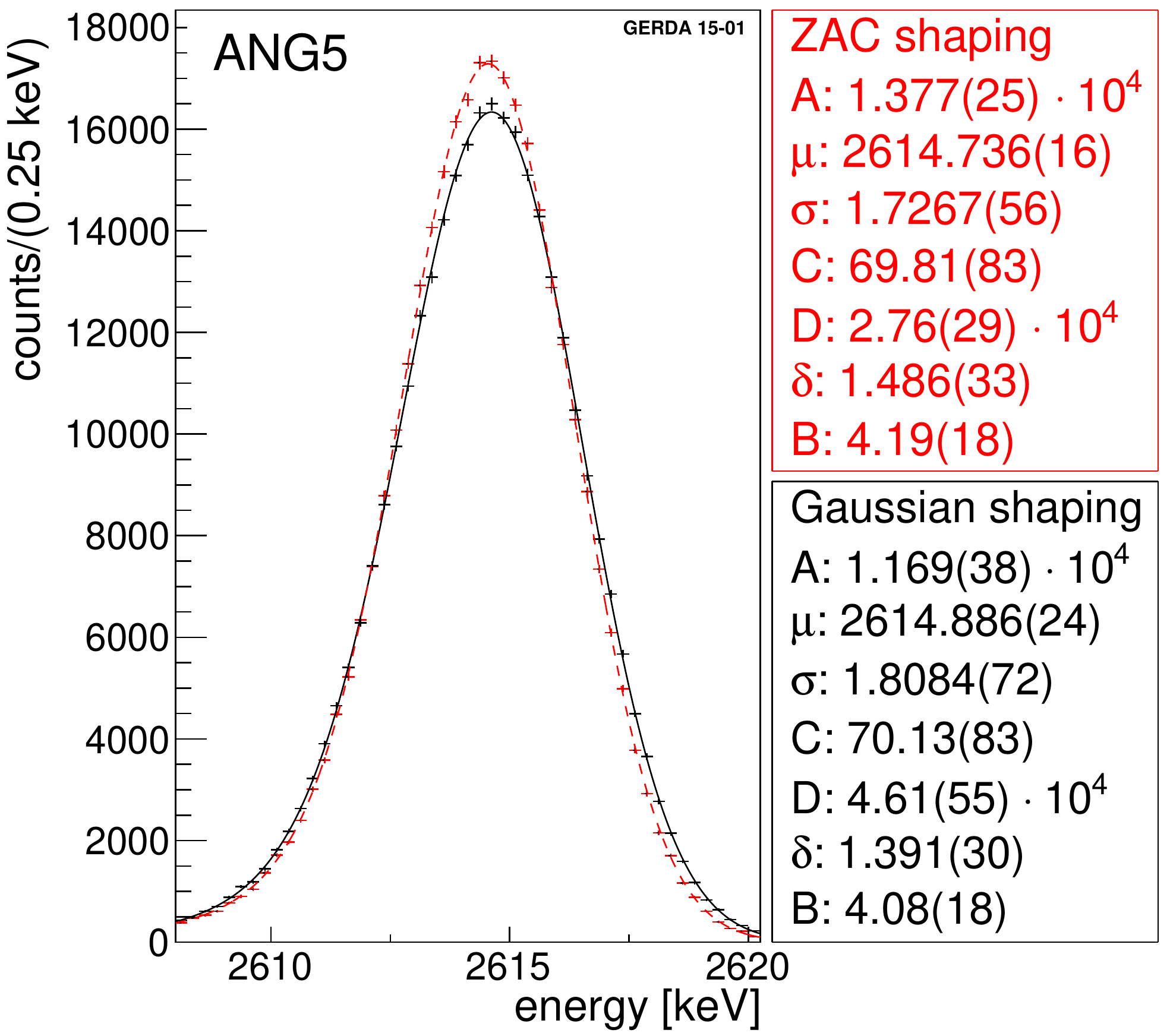}
  \caption{      \label{fig:208Tl_ANG5}
               $^{208}$Tl FEP data for ANG5 at 2614.5~keV.  The curves and
               parameter values corresponding to the best fit for the ZAC and
               the pseudo-Gaussian shaping are shown.
}
\end{center}
\end{figure}
\begin{table}[b]
 \begin{center}
  \caption{  \label{tab:averageFWHM}
           Average FWHM over the complete Phase~I period. The improvement is
           computed as the difference between the FWHM for the pseudo-Gaussian
           and that for the ZAC filter.
           Only the statistical uncertainty due to the peak fit is quoted.
}
  \begin{tabular}{cccc}
    \hline
             & \multicolumn{2}{c}{FWHM at 2614.5~keV[keV]} & improvement \\
    \cline{2-3}
    detector & Gaussian & ZAC & [keV] \\
    \hline
    ANG2  & 4.712(3) & 4.314(3) & 0.398(4) \\
    ANG3  & 4.658(3) & 4.390(3) & 0.268(4) \\
    ANG4  & 4.458(3) & 4.151(3) & 0.307(4) \\
    ANG5  & 4.323(3) & 4.022(3) & 0.301(4) \\
    RG1   & 4.595(4) & 4.365(4) & 0.230(6) \\
    RG2   & 5.036(5) & 4.707(4) & 0.329(6) \\
    GD32B & 2.816(4) & 2.699(3) & 0.117(5) \\
    GD32C & 2.833(3) & 2.702(3) & 0.131(4) \\
    GD32D & 2.959(4) & 2.807(3) & 0.152(5) \\
    GD35B & 3.700(5) & 2.836(3) & 0.864(6) \\
    \hline
  \end{tabular}
 \end{center}
\end{table}

\begin{figure}[t]
\begin{center}
  \includegraphics[width=0.9\columnwidth]{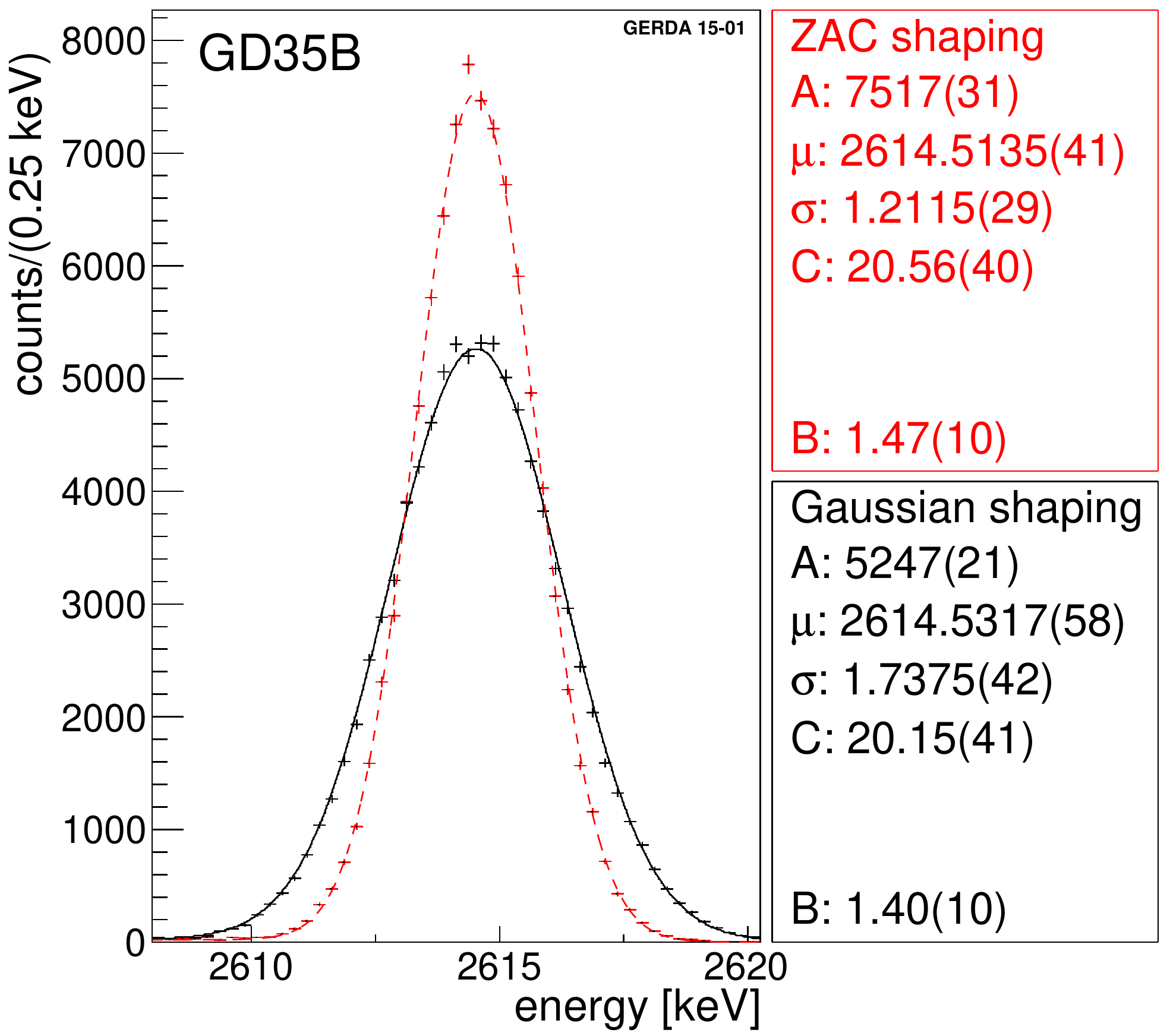}
  \caption{  \label{fig:208Tl_Achilles}
               $^{208}$Tl FEP data for GD35B at 2614.5~keV.
               The curves and parameter values corresponding to the best fit
               for the ZAC and the pseudo-Gaussian shaping are shown.
}
\end{center}
\end{figure}

A first remarkable result is the improvement of the energy resolution between
5 and 23\,\% for the $^{208}$Tl FEP at 2614.5~keV of all the Phase~I data.  As
an example Figs.~\ref{fig:208Tl_ANG5} and~\ref{fig:208Tl_Achilles} show the
summed spectrum of all Phase~I calibrations around the 2614.5~keV line for
ANG5 and GD35B, respectively. In both cases, the amplitude of the Gaussian
component is larger for the spectrum obtained with the optimized ZAC filter
and its width is correspondingly reduced.  The parameters $B$~and $C$
describing the continuum below the peak are compatible for the two shaping
filters.

\begin{figure}[b]
\begin{center}
  \includegraphics[width=0.95\columnwidth]{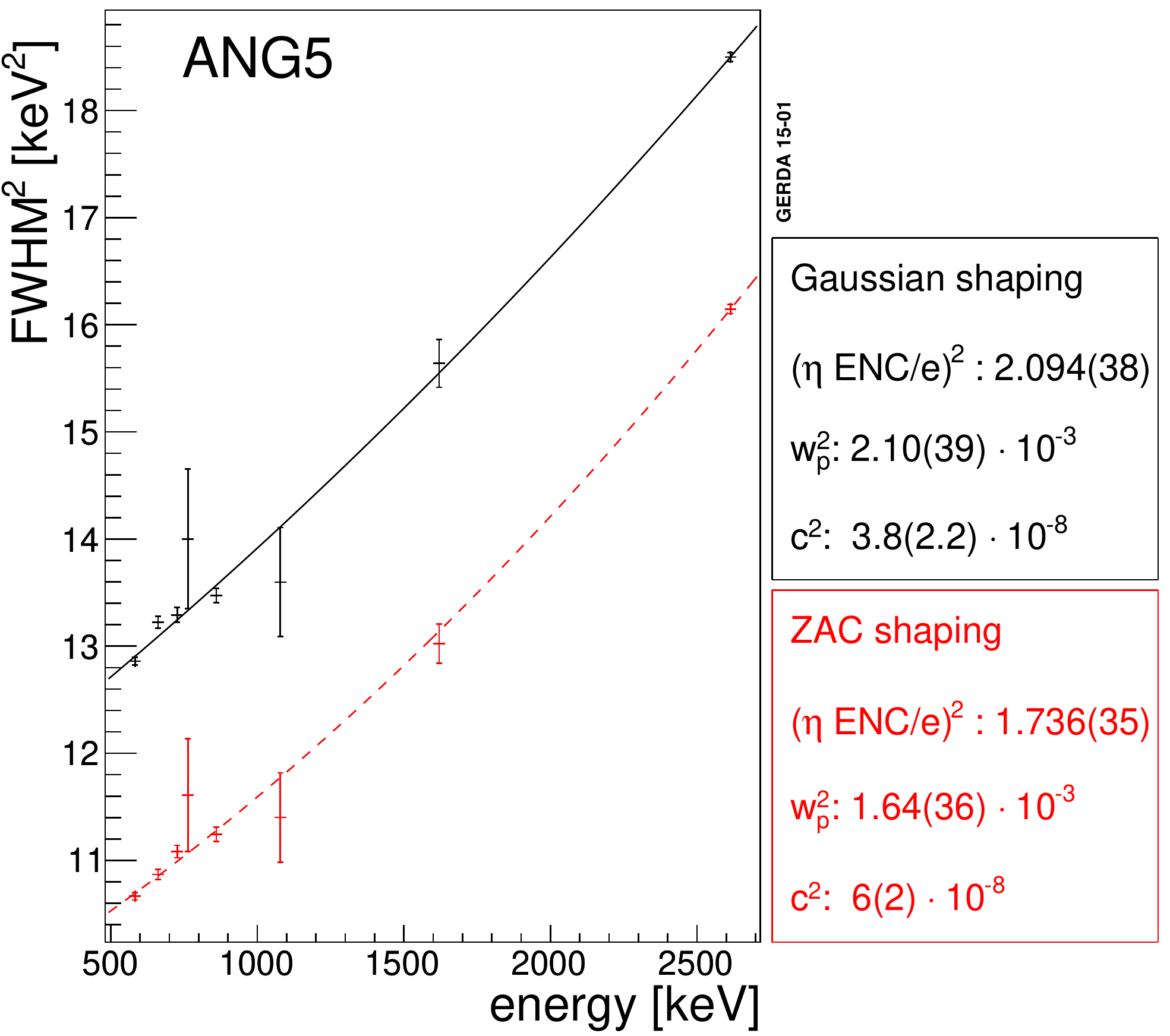}%
  \caption{  \label{fig:FWHMvsEnergy}
             Resolution curve (eq.~\ref{eq:FWHMvsEnergy} with
             $w_p^2=2.355^2\eta F$) for ANG5 calculated for all Phase~I
             calibration spectra merged together for pseudo-Gaussian (full
             line) and ZAC (dashed line) shaping.
}
\end{center}
\end{figure}

\begin{figure*}[t]
\begin{center}
  \includegraphics[width=0.95\textwidth]{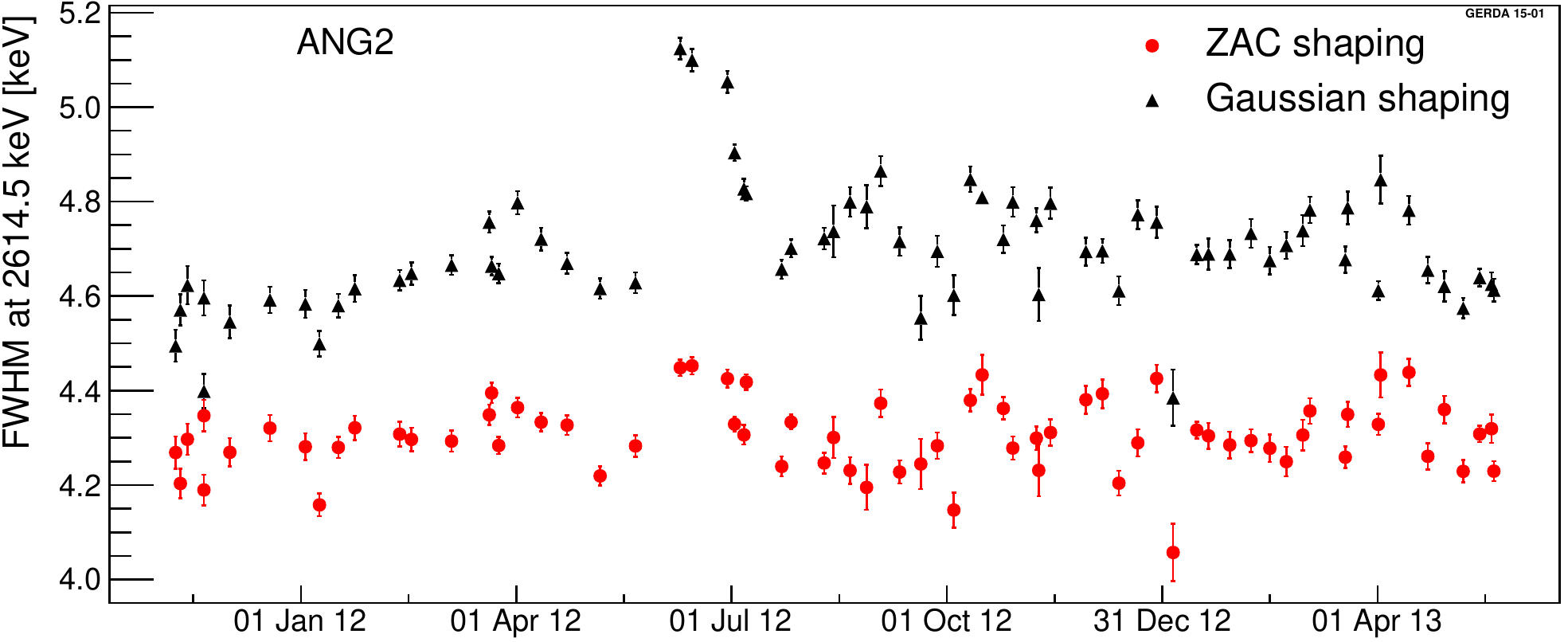}%
  \caption{  \label{fig:FWHMvsTime_ANG2}
                     FWHM of the full energy peak of $^{208}$Tl at 2614.5~keV
                     for ANG2.
}
\end{center}
\end{figure*}

\begin{figure*}[t]
\begin{center}
  \includegraphics[width=0.95\textwidth]{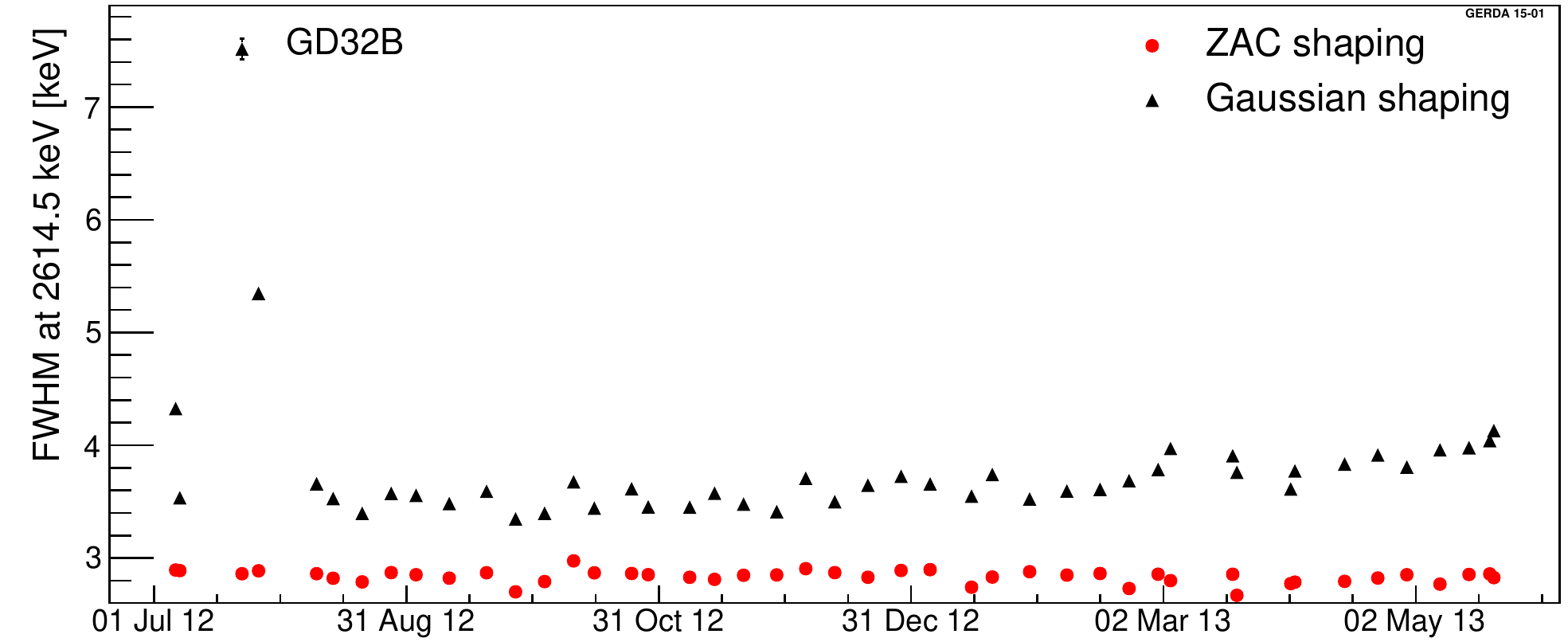}%
  \caption{  \label{fig:FWHMvsTime_Achilles}
                     FWHM of the full energy peak of $^{208}$Tl at 2614.5~keV
                     for GD35B.
                     The error bars are partially within the symbols size.
}
\end{center}
\end{figure*}

While for the coaxial ANG5 a low-energy tail has to be accounted for in the
fit (Fig.~\ref{fig:208Tl_ANG5}) the amplitude of the tail in the BEGe GD35B is
negligible. The tail it therefore automatically removed from the fit
(Fig.~\ref{fig:208Tl_Achilles}). This is attributed to the smaller dimensions
of the BEGe detector and its reduced charge collection inefficiency.  In case
of ANG5 the tail amplitude $D$ is strongly reduced when the ZAC shaping is
used thanks to the presence of the flat-top that allows for an improved
integration of the collected charge.

A deeper understanding of the result is provided by studying the evolution of
the FWHM as function of energy which is fitted according to
Eq.~\ref{eq:FWHMvsEnergy}.  An example is given in Fig.~\ref{fig:FWHMvsEnergy}
showing the resolution curve of all calibration runs for ANG5.  As expected the
major improvement regards the ENC which reduces FWHM$^2$ at all energies by a
constant.  For both, the pseudo-Gaussian and the ZAC filter, the charge
production term $w_p^2=2.355^2\eta F$ is compatible with the theoretical value
of $1.64\cdot10^{-3}$~keV.  Finally, the charge collection term $c^2$ for the
ZAC filter is compatible within the uncertainty with the value obtained for
the pseudo-Gaussian filter.  The large uncertainty of this parameter is due to
the lack of peaks above 3~MeV which makes the fit imprecise.  This term is the
smallest of the three and accounts for maximally 15\,\% of the width at
2614.5~keV.  A consistent behavior is observed for the other detectors as
well.

One of the original motivations for the application of the ZAC filter to the
\gerda\ Phase~I data was the observation of temporary deterioration of the
energy resolution in some detectors interpreted as due to time-evolving
microphonic disturbance not being properly treated by the pseudo-Gaussian
filter.  This is confirmed by the comparison of the FWHM over time for both
filters as shown for ANG2 and GD35B in Figs.~\ref{fig:FWHMvsTime_ANG2}
and~\ref{fig:FWHMvsTime_Achilles}, respectively.  In case of ANG2 the FWHM at
2614.5~keV obtained with the pseudo-Gaussian shaping fluctuates between 4.5
and 4.9~keV.  In June 2012 stronger microphonic disturbance caused a FWHM
increase up to about 5.1~keV.  When using the ZAC filter the effect is
significantly reduced and the FWHM obtained for the affected calibrations is
brought back to a value consistent with the average.  Stronger fluctuations
were present for GD35B.  A very poor energy resolution was observed during the
first month of operation together with a continuous worsening of the
spectroscopic performances in the last four months of Phase~I.  Also in this
case the ZAC filter energy estimate is unaffected by the low-frequency
baseline fluctuations induced by microphonics and allowed to stabilize the
FWHM over time to about 2.8~keV (at 2614.5~keV).

\begin{figure}[t]
\begin{center}
  \includegraphics[width=0.95\columnwidth]{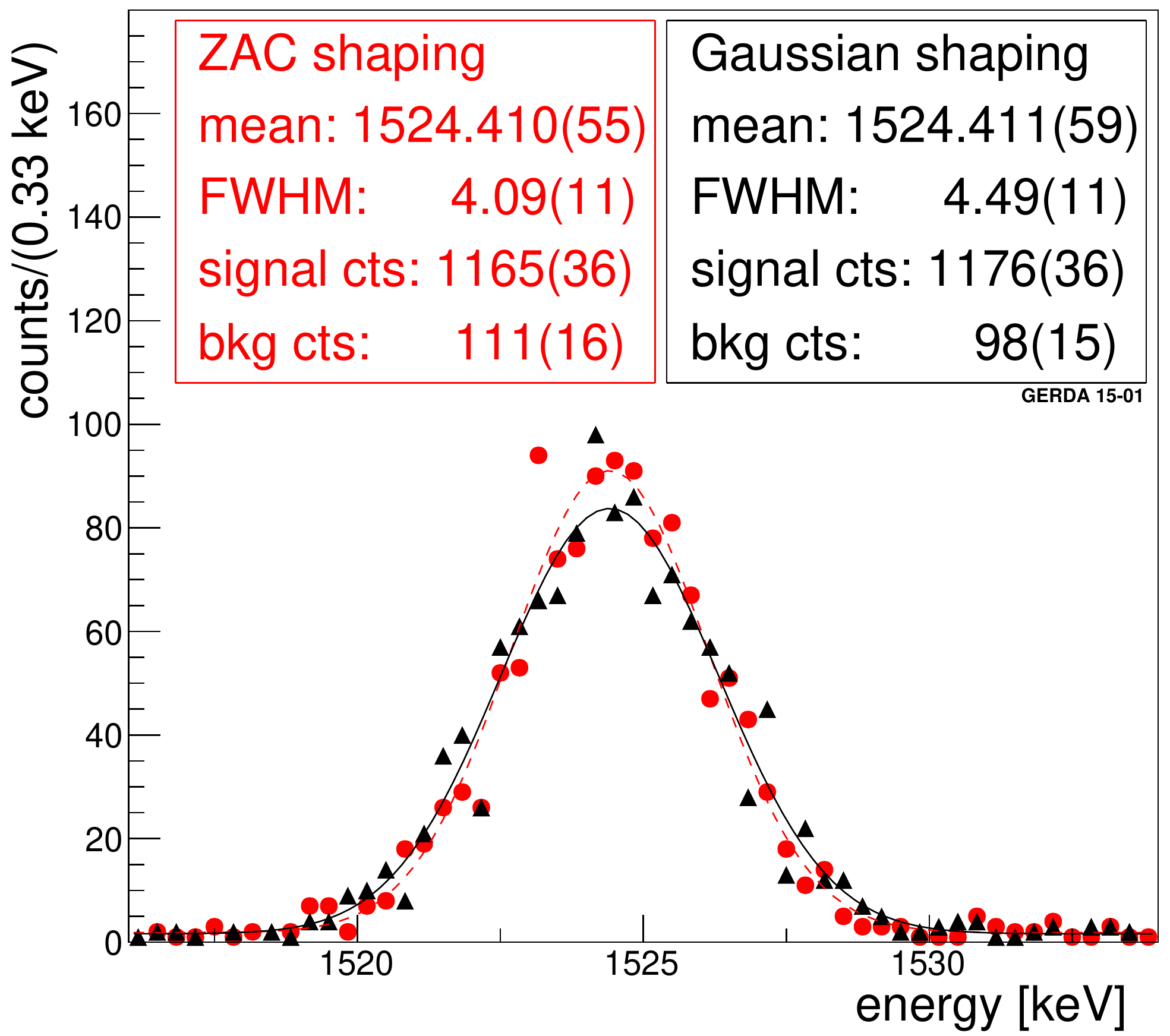}%
  \caption{  \label{fig:CompareGaussCusp_42K_all_sum}
                $^{42}$K peak for the coaxial detectors in the energy spectrum
                for all Phase~I physics runs. The curves and parameter values
                relative to the best fit for the ZAC and the pseudo-Gaussian
                shaping are reported.
}
\end{center}
\end{figure}

The Phase~I average FWHM for the $^{208}$Tl line at 2614.5~keV for each
detector obtained with the pseudo-Gaussian and the ZAC filter are reported in
Table~\ref{tab:averageFWHM}.  The average improvement was calculated as the
difference between the two values.  This is about 0.31~keV for the coaxial and
0.13~keV for the BEGe detectors apart from GD35B for which a much larger
improvement is obtained as described above.

\begin{figure}[t]
\begin{center}
  \includegraphics[width=0.95\columnwidth]{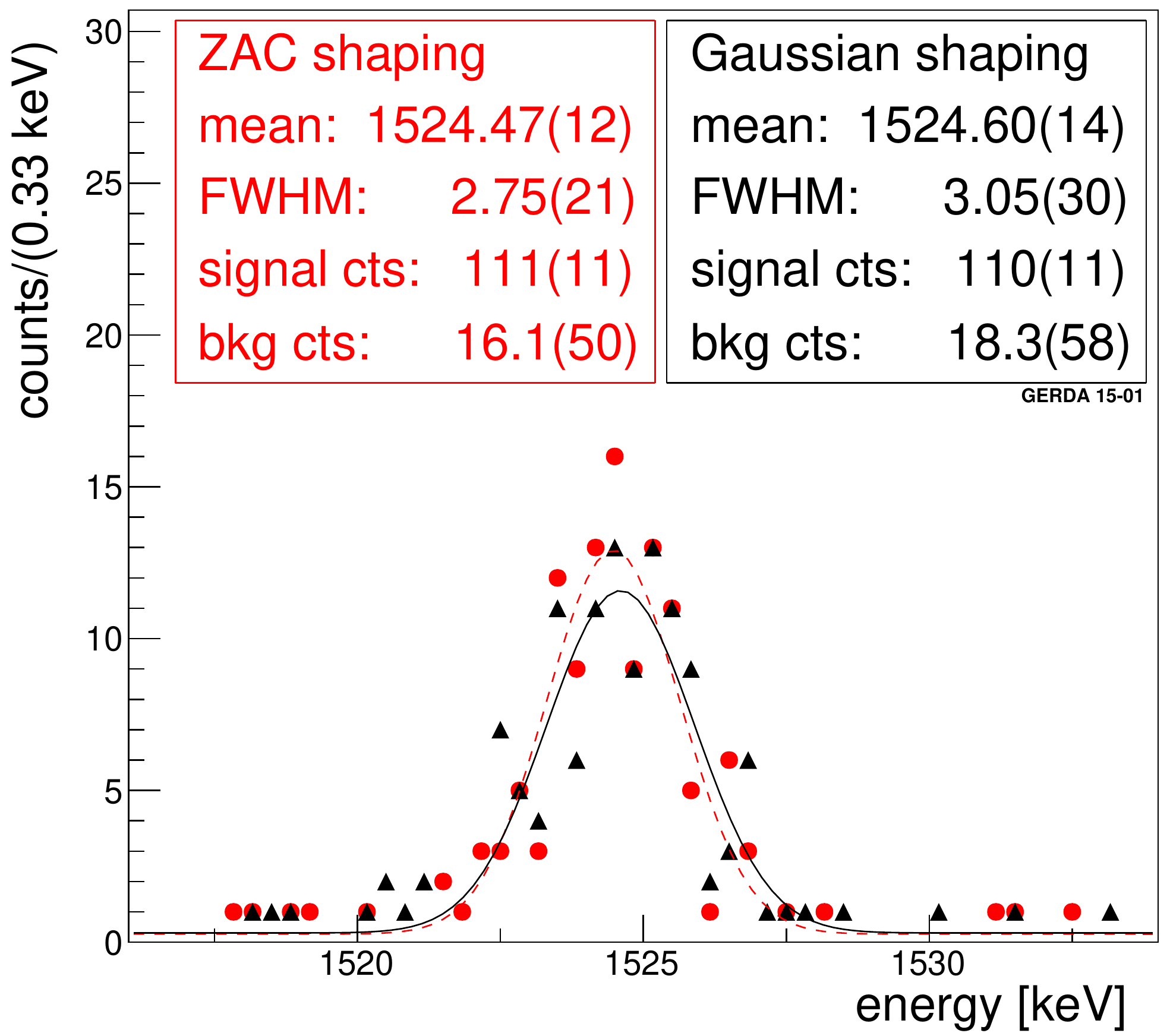}%
  \caption{  \label{fig:CompareGaussCusp_42K_BEGe_sum}
                $^{42}$K peak for the BEGe detectors in the energy spectrum
                for all Phase~I physics runs. The curves and parameter values
                relative to the best fit for the ZAC and the pseudo-Gaussian
                shaping are reported.
}
\end{center}
\end{figure}

The comparison of the effective energy resolution achieved with Phase~I
physics data can be performed exclusively on the $^{42}$K peak at 1524.6~keV
which is the only background line with a sufficient number of counts for a
spectral fit.  The summed energy spectra in the 1515--1535~keV range for all
Phase~I data for the 6 coaxial and the 4 BEGe detectors used for the
\onbb\ decay analysis are shown in Fig.~\ref{fig:CompareGaussCusp_42K_all_sum}
and~\ref{fig:CompareGaussCusp_42K_BEGe_sum}, respectively.  The FWHM obtained
for the pseudo-Gaussian shaping and the coaxial detectors is
$4.49\pm0.11$~keV.  This is 0.30~keV larger than the value expected from the
calibration data.  The reason is given by drifts of the electronics between
calibrations and microphonics
mainly present in ANG2 and ANG4.  For the ZAC filter the drifts
between different physics runs are reduced because the microphonics and the
noise are treated better.  The resulting FWHM of the $^{42}$K peak is
$4.09\pm0.11$~keV and is only $0.15$~keV higher than expected from calibration
data. The net improvement in energy resolution at 1524.6~keV for the coaxial
data is 0.40~keV.  In case of BEGes the ZAC shaping provides a
$2.75\pm0.21$~keV FWHM compared to $3.05\pm0.30$~keV obtained with the
pseudo-Gaussian.  The comparison in this case is harder due to the very
limited number of events.  The improvement on the FWHM of the $^{42}$K line is
in agreement with the expectation from the calibration data.

\begin{figure*}
\begin{center}
  \includegraphics[width=0.95\textwidth]{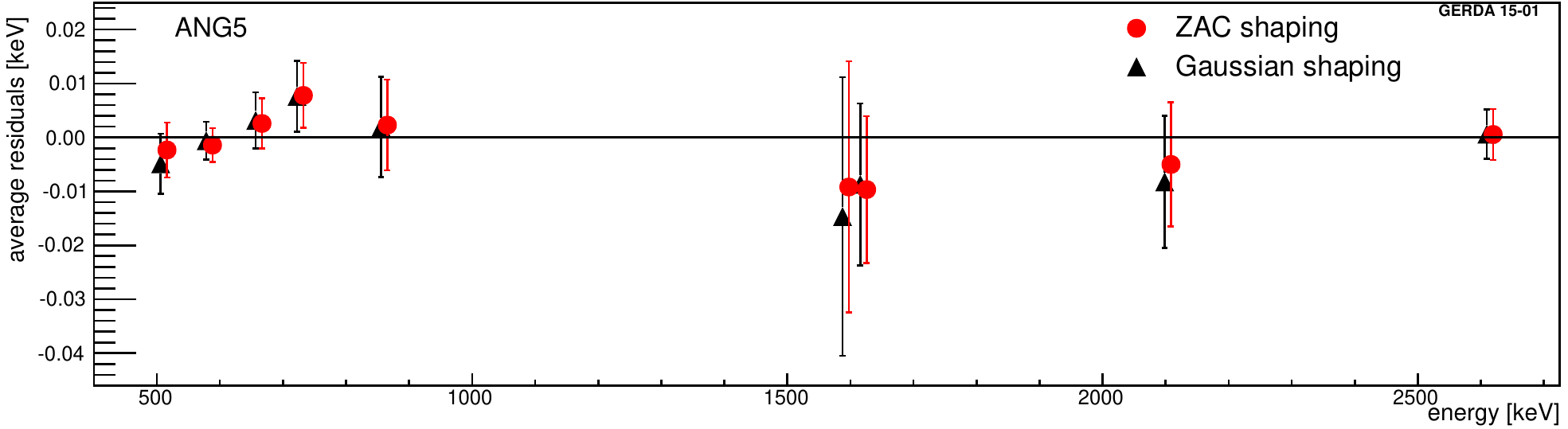}%
  \caption{  \label{fig:averageResiduals_ANG5}
            Average residuals of the $^{228}$Th peak positions relative to
            literature values for ANG5.
            The error bars on the data points correspond to the RMS of the
            residuals for a given peak.
}
\end{center}
\end{figure*}

\begin{figure*}
\begin{center}
  \includegraphics[width=0.95\textwidth]{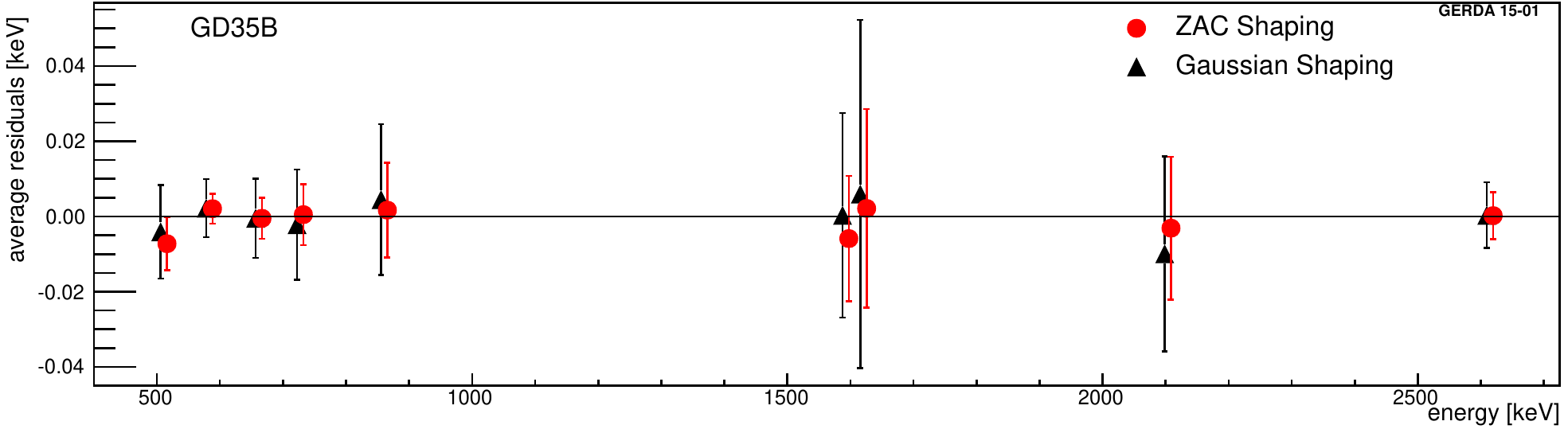}%
  \caption{  \label{fig:averageResiduals_Achilles}
            Average residuals of the $^{228}$Th peak positions relative to
            literature values for GD35B.
            The error bars on the data points correspond to the RMS of the
            residuals for a given peak.
}
\end{center}
\end{figure*}

The improvement in energy resolution given by the ZAC filter is also reflected
in a more precise estimation of the energy scale for the single calibration
runs.  In \gerda\ a second degree polynomial is used as a calibration curve in
order to account for the preamplifier non-linearity.
Figs.~\ref{fig:averageResiduals_ANG5} and~\ref{fig:averageResiduals_Achilles}
show the residuals of the $^{228}$Th peak positions from the corresponding
calibration curve averaged over all Phase~I calibration runs.  Both for the
Gaussian and the ZAC shaping, the average residuals are of order of
$10^{-2}$~keV. Hence, they are much smaller than the peak widths.

A more informative estimation of the energy calibration precision is obtained
by calculating the uncertainty $\delta_{E}$ of the calibration curve at a
given energy, \emph{e.g.} at 1524.6~keV.  For each calibration run the
quantity $\delta_{E}(E=1524.6$~keV) is calculated by error propagation on the
calibration curve parameters.  Using Monte Carlo (MC) simulations
$10^5$~events were randomly generated according to a Gaussian distribution
with zero mean and $\delta_{E}(E=1524.6$~keV). The distributions from all
Phase~I calibration runs are then summed up and the systematic uncertainty of
the energy scale at 1524.6~keV is given by the half-width of the 68\,\%
central interval.  This results to be between 0.03 and 0.07~keV and is up to
$16\%$\ smaller for ZAC shaping with respect to the pseudo-Gaussian filter.

\begin{figure}[tp]
\begin{center}
  \includegraphics[width=0.95\columnwidth]{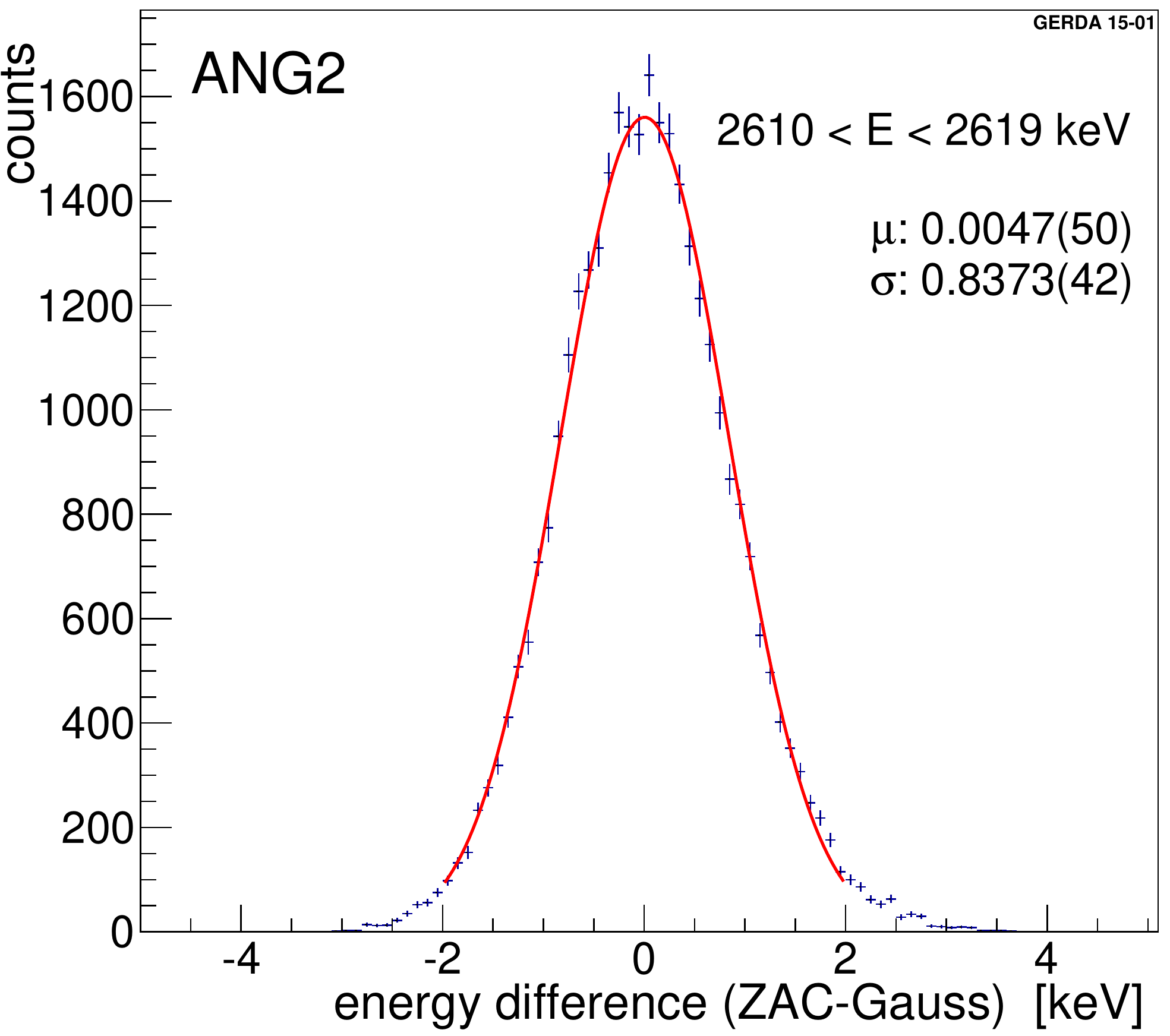}%
  \caption{  \label{fig:DifferenceZACGauss}
         Distribution of the difference between the energy estimated with the
         pseudo-Gaussian and that obtained with the ZAC filter for the
         $^{208}$Tl FEP events at 2614.5~keV. The data refer to a standard
         calibration and are for ANG2.
}
\end{center}
\end{figure}

A cross check of the reprocessed data is given by the event-by-event
comparison of the energy obtained with the ZAC and the pseudo-Gaussian filter.
This is performed by calculating the energy difference of the events in the
2614.5~keV peak as shown in Fig.~\ref{fig:DifferenceZACGauss} for ANG2 during
a typical calibration run.  For all the detectors this distribution is a
Gaussian with a mean value compatible with zero and a width
$\sigma\sim0.8$~keV.  The same behavior is observed at all energies for both
calibration and physics data.

\section{Summary}
\label{sec:summary}

The presence of low-frequency noise in the signals of \gerda\ Phase~I mostly
induced by microphonic disturbance leads to a degraded energy resolution for
some of the deployed detectors.  Spectroscopic performance close to optimal is
obtained by the use of the ZAC shaping filter. This novel Zero Area Cusp
filter is obtained by subtracting two parabolas from the sides of the cusp
filter keeping the area under the parabolas equal to that underlying the
cusp. A selection of calibration runs has been exploited for the optimization
of the ZAC filter. All calibration data sets have then been reprocessed using
the optimal filter parameters.  An average improvement of 0.30~keV in FWHM has
been obtained for both coaxial and BEGe detectors.  In one case (GD35B) the
energy resolution is improved by 0.86~keV with the excellent low-frequency
rejection provided by the ZAC filter.

The stability of the filter parameters over time for the same detector
configuration in \gerda\ along with its outstanding low-frequency noise
rejection capabilities provides a FWHM improvement of 0.40 (0.30) keV at the
$^{42}$K line in the Phase~I physics data for the coaxial (BEGe) detectors.
Any improvement in the energy resolution will increase the sensitivity of the
experiment and allow a better understanding of the experimental background.

The Phase~I physics data, reprocessed with the ZAC shaping, will be combined
with the Phase~II data in a future analysis of the \onbb\ decay.  The
optimization of the shaping filter will be performed from the beginning of
Phase~II following a procedure similar to the one described in the present
work.

\section*{Acknowledgments}

The \gerda\ experiment is supported financially by the German Federal
Ministry for Education and Research (BMBF),
the German Research Foundation (DFG)
via the Excellence Cluster Universe, the Italian Istituto Nazionale
di Fisica Nucleare (INFN), the Max Planck Society (MPG),
the Polish National Science Center (NCN), the Foundation for Polish
Science (MPD programme),
the Russian Foundation for Basic Research (RFBR),
and the Swiss National Science Foundation (SNF).
The institutions acknowledge also internal financial support.

The \GERDA\ collaboration thanks the directors and the staff
of the LNGS for their continuous strong support of the \GERDA\ experiment.

\end{document}

%% file: gerda-abbreviations.tex
\usepackage{amssymb}     
\usepackage{marvosym}    
\usepackage{upgreek}     
\newcommand{\ctsper}      {cts/(keV$\cdot$kg$\cdot$yr)}

\newcommand{\mus}         {{$\upmu$s}}

\newcommand{\qbb}         {{$Q_{\beta\beta}$}}

\newcommand{\onbb}        {{$0\nu\beta\beta$}}

\newcommand{\fgesix}      {\mbox{$f_{76}$}}

\newcommand{\etal}        {\textit{et al.}}

\newcommand{\gerda}       {\textsc{Gerda}}

\newcommand{\GERDA}       {\mbox{\textsc{Gerda}}}  
\newcommand{\hdm}         {\textsc{HdM}}

\newcommand{\gelatio}     {\textsc{Gelatio}}

\newcommand{\gesix}       {{$^{76}$Ge}}

\newcommand{\gess}        {{$^{76}$Ge}}

